\documentclass[%
 twocolumn,
 amsmath,amssymb,
 aps,
]{revtex4}

\usepackage{graphicx}
\usepackage{dcolumn}
\usepackage{bm}

\begin{document}

\title{Electrically charged matter in rigid rotation around magnetized black hole}

\author{Ji\v{r}\'{i} Kov\'{a}\v{r}}
  \email{Jiri.Kovar@fpf.slu.cz}
\author{Petr Slan\'{y}}
\author{Claudio Cremaschini}
\author{Zden\v{e}k Stuchl\'{i}k}%
\affiliation{Institute of Physics, Faculty of Philosophy and Science,
Silesian University in Opava\\ Bezru\v{c}ovo n\'{a}m. 13, CZ-746\,01 Opava, Czech Republic}%

\author{Vladim\'{i}r Karas and Audrey Trova}
\affiliation{Astronomical Institute, Academy of Sciences,
Bo\v{c}n\'{i} II, CZ-141\,31\,Prague, Czech Republic}

\date{\today}

\begin{abstract}
We study charged-fluid toroidal structures surrounding a non-rotating charged black hole immersed in a large-scale, asymptotically uniform  magnetic field. In continuation of our former study on electrically charged matter in approximation of zero conductivity, we demonstrate existence of orbiting structures in permanent rigid rotation in the equatorial plane, and charged clouds hovering near the symmetry axis. We constrain the range of parameters that allow stable configurations and derive the geometrical shape of equi-pressure surfaces. Our simplified analytical study suggests that these regions of stability may be relevant for trapping electrically charged particles and dust grains in some areas of the black hole magnetosphere, being thus important in some astrophysical situations.
\end{abstract}


\maketitle


\section{Introduction}
Equilibrium configurations of perfect fluid play an important role in studies of geometrically thick accretion discs around compact objects and black holes in astrophysics \cite{Fra-Kin-Rai:2002:AccretionPower:,Koz-Jar-Abr:1978:ASTRA:,Abr-Jar-Sik:1978:ASTRA:,Stu-Sla-Hle:2000:ASTRA:,Fon-Dai:2002:MNRAS:,Rez-Zan-Fon:2003:ASTRA:,Stu:2005:MODPLA:,Sla-Stu:2005:CLAQG:,Stu-Sla-Kov:2009:CLAQG:,Kuc-Sla-Stu:2011:CASTRP:}. Physical properties of the system are determined by an interplay of different factors which define the form of the fluid flow; they are, in particular, the gravitational and electromagnetic action of the central body and of the ambient fields. Effects of general relativity are particularly important and visible near horizon of a black hole or neutron star, nevertheless, even a mildly charged matter is strongly influenced by external magnetic fields to which the black hole may be embedded \cite{Kov-Stu-Kar:2008,Kov-Kop-Kar-Stu:2010,Kop-Kar-Kov-Stu:2010,Stu-Kol:2012,Stu-Kol:2014,Tur-Kol:2013}.

Astrophysical fluids can be embedded in different conditions, and so they correspondingly exhibit vastly different behavior. Cosmic plasma is 
often fully ionized, in which case it is described by perfect (infinitely large) conductivity. On the other hand, in a cold environment (e.g. `dead zone' of accretion discs), the dusty medium with imperfect (low) electrical conductivity exists.

Electric charges can develop by several mechanisms, namely, irradiation of dust grains from the central source \cite{Wei:2006}, charge 
exchange within complex plasmas \cite{Dra:1979,Men:1994}, and charge separation by the organised magnetic field \cite{Neu:1993,Pet-etal:2002,Cre-Stu:2014}. Large-scale organised magnetic fields occur, for example, in the vicinity of magnetic stars, pulsars and magnetars, where the surface magnitude of the magnetic induction are dominated by the dipole-type structure and can reach extreme values higher than $\simeq10^{15}\,{\rm G}$ \cite{Tie:2013}.

Although the effects of imperfect (finite) conductivity as well as electric charge separation must occur to certain extent in realistic 
astrophysical environments, these are complications that make the problem intractable to analytical calculations. Here we want to address 
this kind of issues within the frame of idealised (toy-model) approach, which allows us to investigate some fundamental aspects of the problem. Among the main approximations adopted in this paper, we assume non-gravitating (test) fluid immersed in the gravitational field of a static black hole and the uniform magnetic field generated by a distant source; `test' components do not contribute to the spacetime metric. Also, the  electromagnetic field generated by the fluid is neglected. We consider organized rotational motion of the fluid and neglect effects of turbulence. We also set conductivity of the medium to zero, so that charges adher to the circling matter and are carried convectively. Such kind of simplification could be relevant for the hyperdense accretion tori created in the final stages of coalescence of binary systems of neutron star forming a central black hole and accretion torus \cite{Rez:2013}. 

In our recent paper \cite{Kov-etal:2011}, we proceeded by using a number of additional simplifications that allowed us to clearly illustrate the 
model. Namely, we assumed the gravitational and electromagnetic fields combined within the Reissner-Nordstr{\o}m solution of the Einstein-Maxwell 
equations \cite{MTW}. For the fluid we assumed the polytropic equation of state fitting for analytical calculations, constant profile of the specific angular-momentum throughout the torus, and constant specific charge everywhere in the equatorial plane. Thanks to these simplifications, we were able to formulate and solve a non-trivial, theoretically reasonable system in an analytical framework.

The presence of electric charge of the Reissner-Nordstr{\o}m black hole makes this solution unlikely from the astrophysical point of view 
because its global charge is thought to be quickly neutralized by selective accretion of opposite charges from the surrounding medium although a small value of a net equilibrium charge is expected. On the other hand, large-scale magnetic fields do exist in the Universe, both on stellar and galactic scales, and cosmic black holes must interact with them. Hence, it is relevant to discuss our problem in the context of magnetized black 
holes. To avoid tedious calculations, we start with the weak-field solution of asymptotically homogeneous magnetic field \cite{Wal:1974}. We adopt the polytropic equation of state as in the previous work, however, we allow for a general value of the polytropic exponent \cite{Avinash2006}. 
Furthermore, we set the constant angular velocity throughout the torus (rigid rotation) that could be relevant for slender tori or hyperdense tori. We find it interesting that the above-mentioned restrictions lead to a solution (which we give in an analytical form), even if astrophysically realistic tori will very likely establish some kind of a more complicated distribution of the velocity field and the corresponding angular momentum. Apart from the stable solution of toroidal shape located near the equatorial plane, we have found also possible existence of spheroidal clouds located along the polar axis. Clearly, the location of the tori and clouds depends on the specific charge of the fluid, charge of the black hole and the strength of the magnetic field of the external origin. Both the kinds of structure are astrophysically relevant. The equatorial tori can represent accretion discs, whereas the polar clouds can scatter and polarize light on the axis, as reported on in some objects \cite{Ant:1993,Mar:2014}.

Throughout the paper, we use the geometrical system of units for quantities denoted by a bar, $\bar{x}$. These become dimensionless $x$ when 
scaled by the the mass of central black hole. For the direct interpretation, we express our results also in physical (SI) units, 
denoted as $\hat{x}$.
\section{Charged structures in Wald field}
We built our model of an electrically charged perfect fluid torus in an axially symmetric spacetime endowed with an electromagnetic field of the same symmetry. In the spherical polar coordinates $(t,\phi,r,\theta)$, the considered background is described by an axially symmetric metric tensor $g_{\alpha\beta}$ and by an electromagnetic field tensor $F_{\alpha\beta}=\nabla_{\alpha}A_{\beta}-\nabla_{\beta}A_{\alpha}$, where, due to the symmetry, the vector potential has the form $A_{\alpha}=(A_t,A_{\phi},0,0)$. To this end, we adopt here the approximation of electromagnetic test field in the curved spacetime of a black hole. 

\subsection{Pressure equations}
As the main assumption of the model, the perfect fluid is characterized by its rotation in the \mbox{$\phi$-direction} only, with the \mbox{4-velocity} $U^{\alpha}=(U^t,U^{\phi},0,0)$, specific angular momentum $\ell=-U_{\phi}/U_t$ and angular velocity (with respect to 
distant observers) $\omega=U^{\phi}/U^t$, related by the formulae
\begin{eqnarray}
\label{Omega}
\omega&=&-\frac{\ell g_{tt}+g_{t\phi}}{\ell g_{t\phi}+g_{\phi\phi}},\\
(U_t)^2&=&\frac{g_{t\phi}^2-g_{tt}g_{\phi\phi}}{\ell^2 g_{tt}+2\ell g_{t\phi}+g_{\phi\phi}}.
\end{eqnarray}
The shape of the rotating charged fluid with a charge density profile $q$ and energy density $\epsilon$ is determined by isobaric surfaces of the isotropic pressure $p$ profile (equi-pressure surfaces), which can be determined from two partial differential equations 
\begin{eqnarray}
\label{pressure}
\partial_r p&=&-(p+\epsilon)\mathcal{R}_1+q\mathcal{R}_2\equiv \mathcal{R}_0,\nonumber\\
\partial_{\theta} p&=&-(p+\epsilon)\mathcal{T}_1+q\mathcal{T}_2\equiv \mathcal{T}_0,
\end{eqnarray}
where $\mathcal{R}_0=\mathcal{R}_0(r,\theta)$ and $\mathcal{T}_0=\mathcal{T}_0(r,\theta)$ denote the right hand sides of these equations, and
\begin{eqnarray}
\mathcal{R}_1&=&\partial_r\,\ln{|U_t|} - \frac{\omega \partial_r \ell}{1-\omega \ell},\\
\mathcal{R}_2&=&U^t\partial_r A_t+U^{\phi}\partial_r A_{\phi},\\
\mathcal{T}_1&=&\partial_{\theta}\,\ln{|U_t|} - \frac{\omega \partial_{\theta} \ell}{1-\omega \ell},\\
\mathcal{T}_2&=&U^t\partial_{\theta} A_t+U^{\phi}\partial_{\theta} A_{\phi}.
\end{eqnarray}
These pressure equations follow from the conservation laws and Maxwell’s equations and we describe their derivation in details in Appendix.

Besides the pressure $p$ and energy density $\epsilon$, the other fluid variable considered here is the rest-mass density $\rho$. The thermodynamical description of the fluid is then specified by choosing an appropriate equation of state $p=p(\epsilon,q)$, which also involves the charge density of the fluid, describing the contribution of the Coulomb interaction between the fluid particles to the total pressure.
\subsection{Equation of state and angular momentum}
The pressure and density profiles designating shapes of the tori can be determined from the model, providing either the specific angular momentum or charge density through the torus are specified. An equation of state must be also fixed in order to close the set of equations. 

In general, the pressure equations (\ref{pressure}) are not integrable. In addition, the integrability condition
\begin{eqnarray}
\label{icondition}
\partial_{\theta} \mathcal{R}_0+\mathcal{T}_0\partial_p \mathcal{R}_0 =\partial_r \mathcal{T}_0+\mathcal{R}_0\partial_p \mathcal{T}_0
\end{eqnarray}
must be also satisfied. The existence of a solution of equations (\ref{pressure}) here is guaranteed for zero charged density $q=0$ when the last terms in 
equations (\ref{pressure}) vanish and we get the Euler equation describing a rotating uncharged perfect fluid (see, e.g., papers \cite{Koz-Jar-Abr:1978:ASTRA:,Abr-Jar-Sik:1978:ASTRA:}).
In our case when $q\neq 0$, the situation is more complicated and equations (\ref{pressure}) are no longer integrable for arbitrary $q=q(r,\theta)$ and arbitrarily chosen $\ell=\ell(r,\theta)$. Thus, it is necessary either to specify $\ell=\ell(r,\theta)$ (with even $\ell={\rm const}$ being possible) and find an appropriate $q=q(r,\theta)$, which is consistent with that or, vice versa, to specify $q=q(r,\theta)$ (with even $q={\rm const}$ being possible) and find an appropriate $\ell=\ell(r,\theta)$. This is, however, strictly `necessary' only if the equation of state is prescribed. Otherwise, one could also absorb the constraint into the equation of state. 

The charged tori must clearly have distributions of charge and angular momentum satisfying the integrability condition. 
In this paper, we assume unique configuration represented by the rigidly rotating matter $\omega={\rm const}$ throughout the torus. This conception supports the idea of the fluid having bulk hydrodynamic motion with no friction, predominating over the electromagnetic effects, which the fluid with almost infinite electric resistivity must follow. The corresponding specific charge distribution is then determined from the integrability condition.

For an uncharged perfect fluid, the polytropic relation 
\begin{eqnarray}
\label{EOS}
p=\kappa \rho^{\Gamma},
\end{eqnarray}
with $\kappa$ and $\Gamma$ being a polytropic coefficient and exponent, is very often chosen as the equation of state. This is widely-used simple relation, which embodies conservation of entropy, appropriate for a perfect fluid. However, assuming high values of $\kappa$ and high temperatures of the fluid, when the electrostatic corrections to the equation of state become negligible, we can use this polytropic equation of state consistently even in our charged case. Moreover, we choose values of $\rho$ sufficiently low so that the medium is non-relativistic and the contribution of the specific internal energy $\epsilon/\rho-1$ to the total energy density is then negligible, i.e., $\epsilon \approx \rho$. As it is shown below, this approximation is consistent also with the assumption of negligible self-electromagnetic field.
\subsection{Pressure equations transformation}
Having chosen the polytropic equation of state and fixed specific angular momentum due to the $\omega={\rm const}$ distribution, it is possible to determine the charge density distribution from the integrability condition (\ref{icondition}) and to calculate the final pressure profile from pressure equations (\ref{pressure}). In general, in dependence on the chosen compact object background, and parameters of the matter, it might be a difficult task, providing us with a numerical solution only. Alternatively, it is more convenient to assume an equivalent set of differential equations for the density $\rho$, related to the pressure through the equation of state; the charge density distribution can be also written in the natural form $q=\rho q_{\rm s}$, where $q_{\rm s}=q_{\rm s}(r,\theta)$ is now an unknown function to be determined from the integrability condition, corresponding to the specific charge distribution. We applied this approach in the Reissner-Nordstr\o m case in the paper \cite{Kov-etal:2011}, being motivated by the simplification of the obtained non-linear differential equations in the chosen specific case of polytropic exponent $\Gamma=2$. 

Here, to get the analytical expression of final pressure profiles for a general polytropic exponent, we introduce a transformation, allowing us to work with linear differential equations instead of the non-linear pressure ones (\ref{pressure}). At first, it is convenient to define the function 
\begin{eqnarray}
K=\frac{q}{\rho+p},
\end{eqnarray}
which implies the set of equations
\begin{eqnarray}
\label{pressure1}
\partial_r p&=&-(p+\rho)\mathcal{R},\nonumber\\
\partial_{\theta} p&=&-(p+\rho)\mathcal{T},
\end{eqnarray}
where $\mathcal{R}=\mathcal{R}_1-K\mathcal{R}_2$ and $\mathcal{T}=\mathcal{T}_1-K\mathcal{T}_2$. Introducing the auxiliary function
\begin{eqnarray}
\label{transf}
h&=&\ln(1+\kappa\rho^{\Gamma-1})=\ln\left(1+\frac{p}{\rho}\right),
\end{eqnarray}
we get the set of linear equations
\begin{eqnarray}
\label{pressure2}
\partial_r h&=&-\frac{\Gamma-1}{\Gamma}\mathcal{R},\nonumber\\
\partial_{\theta} h&=&-\frac{\Gamma-1}{\Gamma}\mathcal{T},
\end{eqnarray}
whereas the integrability condition for them reads 
\begin{eqnarray}
\label{condition1}
\partial_{\theta} \mathcal{R}=\partial_r \mathcal{T}.
\end{eqnarray}
The function $K$ is addressed as the `correction' function, ensuring the integrability of the equations (\ref{pressure2}) and determining the specific charge distribution.
\subsection{Gravity and electromagnetic test-field solution}
We consider our tori rotating close to a static charged black hole immersed in an asymptotically homogeneous magnetic field (see Fig. \ref{Fig:1}). Such a setting can be described by the special case (zero rotational parameter) of Wald's test-field solution of Einstein-Maxwell equations \cite{Wal:1974}, which in dimensionless units reads 
\begin{eqnarray}
\label{wald}
A_t=-\frac{Q}{r},\quad
A_{\phi}=\textstyle{\frac{1}{2}}\,Br^2\sin^2{\theta},
\end{eqnarray}
describing the electromagnetic field in the background of Schwarzschild geometry 
\begin{eqnarray}
\label{metric}
{\rm d}s^2&=&-\left(1-\frac{2}{r}\right){\rm d}t^2+\left(1-\frac{2}{r}\right)^{-1}{\rm d}r^2\\\nonumber
&&+r^2({\rm d}\theta^2+\sin^2{\theta}{\rm d}\phi^2).
\end{eqnarray}
Note that the parameters $Q$ and $B$ are only test-field parameters and do not influence the spacetime geometry.
The parameter $Q$ represents the dimensionless charge related to the black hole and the $t$-component of the vector potential can be thus identified with the component of the vector potential of the Reisnerr-Nordstr\o m solution \cite{MTW}. The parameter $B$ corresponds to the dimensionless strength of the uniform magnetic field, as it can be easily checked from the asymptotic behavior of the $\phi$ component of the vector potential. For completeness, we can mention that the full Wald's test field solution describes the homogeneous magnetic field in the background of not the only static Schwarzschild black hole, but the rotating Kerr black hole. 

Let us note that a generalization of the adopted approach to the case of a rotating (Kerr) black hole is straightforward in principle, but it would technically be very tedious because of the metric coefficient $g_{t\phi}$ governing the frame dragging. 
\subsection{Correction function}
In the assumed Schwarzschild-Wald background, the integrability condition (\ref{condition1}) implies a partial differential equation of the first order for the correction function. Its analytical solution has the form 
\begin{eqnarray}
\label{correction}
K=\sqrt{\mathcal{P}}\;\mathcal{C}_{\rm i}(r,\theta;Q,B,\omega),
\end{eqnarray}
where
\begin{eqnarray}
\mathcal{P}=1-\frac{2}{r}-r^2\omega^2\sin^2{\theta}.
\end{eqnarray} 
The function $\mathcal{C}_{\rm i}(r,\theta;Q,B,\omega)$ is a generic function from the class of equivalent functions 
\begin{eqnarray}
\label{generic1}
\mathcal{C}_1&=&\mathcal{C}_1\left[\frac{2Q}{r}-Br^2\omega\sin^2{\theta}\right],\\
\label{generic2}
\mathcal{C}_2&=&\mathcal{C}_2\left[\frac{2}{r}-\frac{B}{Q}r^2\omega\sin^2{\theta}\right],\\
\label{generic3}
\mathcal{C}_3&=&\mathcal{C}_3\left[\frac{2Q}{Br}-r^2\omega\sin^2{\theta}\right],\\
\label{generic4}
\mathcal{C}_4&=&\mathcal{C}_4\left[\frac{2Q}{B\omega r}-r^2\sin^2{\theta}\right],\\
{\rm etc.} 
\end{eqnarray}
which must be chosen to specify the full solution for the correction function.  Actually, it is the way how to specify the specific charge distribution in our approach. In two limiting cases ($Q=0$, $B\neq 0$) and ($B=0$, $Q\neq 0$), we get the two generic functions $\mathcal{C}_{\rm B}=\mathcal{C}_{\rm B}[r\sin{\theta}]$ and $\mathcal{C}_{\rm Q}=\mathcal{C}_{\rm Q}[r]$, respectively, as it is obvious from, e.g., the generic  function form (\ref{generic1}). 

\subsection{Equatorial tori and polar clouds existence}
\begin{figure}
\centering
\includegraphics[width=1\hsize]{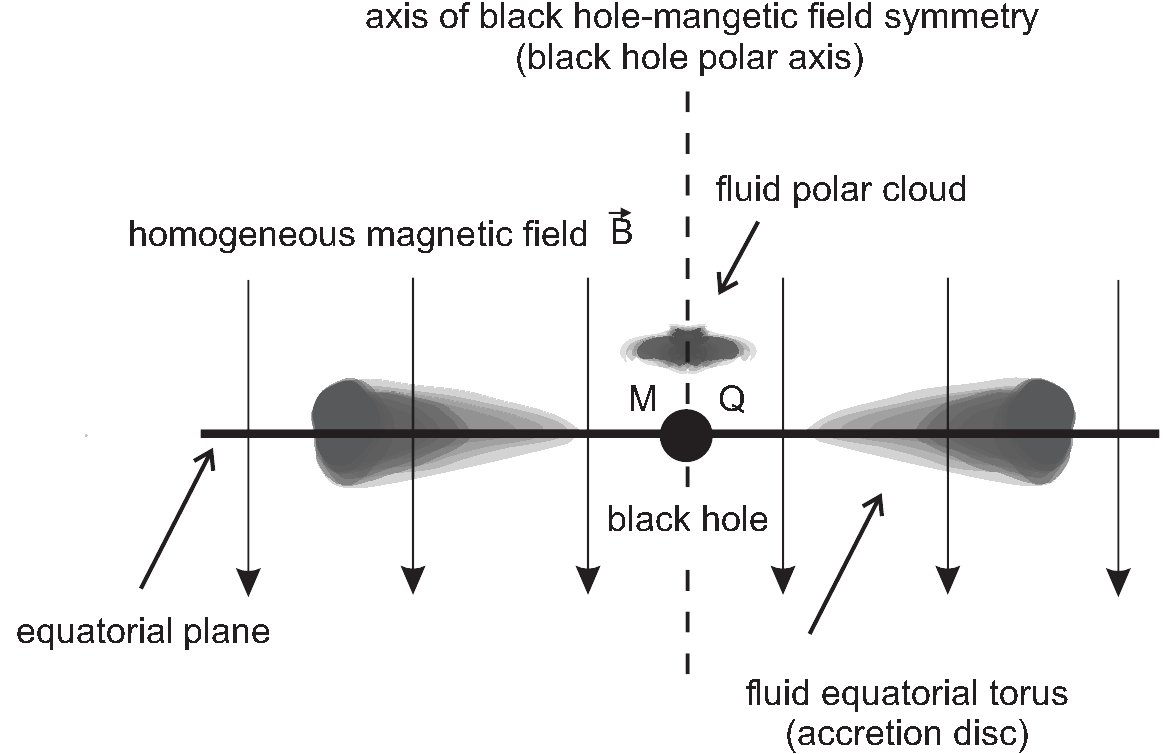}
\caption{Illustration of black hole with mass $M$ and charge $Q$ immersed in asymptotically homogeneous magnetic field $\vec{B}$ encircled by a torus-like equatorial configuration and a polar cloud.}
\label{Fig:1}
\end{figure}
In this work, we are interested in the existence of the rigidly rotating tori centered in the equatorial plane and also in the structures centered on the axis of symmetry referred to as polar clouds (see figure \ref{Fig:1}). As stated above, the shape of tori (clouds) is well determined by the equi-pressure ($p={\rm const}$) or equivalently by equi-density ($\rho={\rm const}$) surfaces and, as we can now see from  relation (\ref{transf}), also by the surfaces of constant values of the function $h={\rm const}$; the boundary of the torus corresponds to the zero-surface of $h$ and consequently to the zero-surfaces of $p$ and $\rho$ . 

Having the pressure equations (\ref{pressure}) or the alternative equations (\ref{pressure1}) or (\ref{pressure2}) available, we can, however, investigate the tori existence even without their final solution.
We denote the center of the torus $(r_{\rm c},\theta_{\rm c})$ as the point where the function $h$ (pressure, density) takes its maximal value, i.e., where the necessary conditions 
\begin{eqnarray}
\label{necessary}
\partial_r h|_{(r=r_{\rm c},\theta=\theta_{\rm c})}=0,\quad \partial_{\theta} h|_{(r=r_{\rm c},\theta=\theta_{\rm c})}=0
\end{eqnarray}
or equivalently $\mathcal{R}_{(r=r_{\rm c},\theta=\theta_{\rm c})}=0$ and $\mathcal{T}_{(r=r_{\rm c},\theta=\theta_{\rm c})}=0$ are satisfied. 
These conditions, however, are only the necessary conditions for an existence of local extrema. To verify the real existence of the tori, we have to construct the Hessian matrix
\begin{eqnarray}                                                              
\label{Hessian}
\mathcal{H} =
\left( 
\begin{array}[c]{cc}
\partial^2_{rr} h & \partial^2_{r\theta} h\\
\partial^2_{\theta r} h & \partial^2_{\theta\theta} h
\end{array}
\right),
\end{eqnarray}
whereas in the loci of maxima, two conditions 
\begin{eqnarray}
\label{sufficient1}                                                                  
\partial^2_{rr} h|_{(r=r_{\rm c},\theta=\theta_{\rm c})}<0,\quad
\det\mathcal{H}|_{(r=r_{\rm c},\theta=\theta_{\rm c})}>0
\end{eqnarray}
must be satisfied. Because of the background and tori symmetry, the mixed partial derivatives in the equatorial plane and on the axis of symmetry vanish and the conditions for maxima reduce to 
\begin{eqnarray}
\label{sufficient2}                                                                  
\partial^2_{rr} h|_{(r=r_{\rm c},\theta=\theta_{\rm c})}<0,\quad
\partial^2_{\theta\theta} h|_{(r=r_{\rm c},\theta=\theta_{\rm c})}<0.
\end{eqnarray}
Finally, we have to realize that the correction function (\ref{correction}) limits the existence of the tori center only to the region where 
\begin{eqnarray}
r_{\rm c}-2-r_{\rm c}^3\omega^2\sin^2\theta_{\rm c}>0.
\end{eqnarray}
Being interested in the existence of the toroidal structures, we first consider the two limiting cases ($Q=0$, $B\neq 0$) and ($B=0$, $Q\neq 0$) separately, then we consider the general situation ($Q\neq 0$, $B\neq 0$).  

In the $Q=0$ case, the first of necessary conditions (\ref{necessary}) implies the stationary (possible maxima) points   satisfying the relation 
\begin{eqnarray}
B=\frac{1-\omega^2 r_{\rm c}^3\sin^2{\theta}_{\rm c}}{r_{\rm c}^2\sin^2{\theta}_{\rm c}\omega\mathcal{C}_{\rm B}(r_{\rm c}-2-\omega^2r_{\rm c}^3\sin^2{\theta}_{\rm c})},
\end{eqnarray}
whereas the second one is automatically satisfied for $\theta_{\rm c}=0$ and $\theta_{\rm c}=\pi/2$. In the case of polar clouds, where $\theta_{\rm c}=0$, the first sufficient condition (\ref{sufficient2}) implies $r_{\rm c}<3/4$, thus there are no polar clouds possible in this special case, since we are restricted to the stationary part of the spacetime ($r>2$) above the black hole horizon only. On the other hand, the equatorial tori, where $\theta_{\rm c}=\pi/2$, can exist. The second sufficient condition (\ref{sufficient2}) reads $r_{\rm c}-2-r_{\rm c}^3\omega^2>0$, and can be easily satisfied for sufficiently low values of $\omega$. The first sufficient condition implies
\begin{eqnarray}
\frac{(3r_{\rm c}-4-7r_{\rm c}^3\omega^2+2r_{\rm c}^6\omega^4)}{r_{\rm c}(r_{\rm c}-2-r_{\rm c}^3\omega^2)}<(r_{\rm c}^3\omega^2-1)\frac{\partial_r\mathcal{C}^{\pi/2}_{\rm B}}{\mathcal{C}^{\pi/2}_{\rm B}},
\end{eqnarray}
which can be also satisfied for conveniently chosen profile of the generic function $\mathcal{C}^{\pi/2}_{\rm B}[r]=\mathcal{C}_{\rm B}|_{\theta=\pi/2}$, being the function of $r$ in the equatorial plane only. 

In the $B=0$ case, as it follows from the first of necessary conditions (\ref{necessary}), the stationary points can be determined from relation  
\begin{eqnarray}
Q=\frac{\sqrt{r_{\rm c}}(1-\omega^2 r_{\rm c}^3\sin{\theta}_{\rm c})}{\sqrt{2}\mathcal{C}_{\rm Q}(r_{\rm c}-2-\omega^2r_{\rm c}^3\sin^2{\theta}_{\rm c})};
\end{eqnarray}
the second from the conditions (\ref{necessary}) here is also automatically satisfied for $\theta_{\rm c}=0$ and $\theta_{\rm c}=\pi/2$. In the case of $\theta_{\rm c}=0$, the first sufficient condition (\ref{sufficient2}) implies $r_{\rm c}<2$, thus there are no polar clouds possible in this special case either. The equatorial tori with $\theta_{\rm c}=\pi/2$ can exist here, as well as in the previous $Q=0$ case. The second sufficient condition (\ref{sufficient2}) reads $r_{\rm c}-2-r_{\rm c}^3\omega^2>0$, and can be met for sufficiently low values of $\omega$. The first sufficient condition implies
\begin{eqnarray}
\frac{2+r_{\rm c}+r_{\rm c}^3(5r_{\rm c}-19)\omega^2-r_{\rm c}^6\omega^4}{2r_{\rm c}(r_{\rm c}-2-r_{\rm c}^3\omega^2)}<(r_{\rm c}^3\omega^2-1)\frac{\partial_r\mathcal{C}^{\pi/2}_{\rm Q}}{\mathcal{C}^{\pi/2}_{\rm Q}},
\end{eqnarray}
which can be also fulfilled for conveniently chosen profile of the generic function $\mathcal{C}^{\pi/2}_{\rm Q}[r]=\mathcal{C}_{\rm Q}[r]$.

The existence of equatorial tori has been proved in both the above mentioned special cases. But we are still interested in the existence of the polar clouds. As we show further, such structures can exist when the influence of the ambient magnetic field is properly combined with the influence of the electric charge of the black hole. We understand that the presence of electrically charged black holes is astrophysically highly theoretical, but admissible due to the influence of the external magnetic field \cite{Wal:1974}. From the theoretical point of view, however, this is very interesting case. Choosing the generic function in the form $\mathcal{C}_2$ (see relation (\ref{generic2})), on the axis of symmetry ($\theta=0$), the correction function reduces to
\begin{eqnarray}
K^{0}=\sqrt{1-u}\;\mathcal{C}^{0}_2(u),
\end{eqnarray}
where $u=2/r$ and $\mathcal{C}^{0}_2(u)=\mathcal{C}_2|_{\theta=0}$. The first of necessary conditions (\ref{necessary}) restricts the stationary points on the axis of symmetry to the loci where  
\begin{eqnarray}
Q=\frac{r_{\rm c}}{\mathcal{C}^{0}_2(r_{\rm c}-2)};
\end{eqnarray} 
the sufficient conditions (\ref{sufficient2}) imply for the stationary points along symmetry axis the conditions
\begin{eqnarray}
\frac{(r_{\rm c}-2)\partial_u \mathcal{C}_2^{0}}{r\mathcal{C}_2^{0}}>1,\\
-\frac{r_{\rm c}\omega^2}{r_{\rm c}-2}>B\omega\mathcal{C}_2^{0},
\end{eqnarray}
which can be satisfied for conveniently chosen profile of the generic function $\mathcal{C}_2$. Thus, the polar clouds really can exist in the general case ($Q\neq 0$, $B\neq 0$).

\subsection{Solution - characteristics profiles}
Adopting the polytropic equation of state (\ref{EOS}), the pressure profiles $p(r,\theta)$ of the tori can be simply obtained through the $h$-function (\ref{transf}) solution. This is determined by the set of differential equations (\ref{pressure2}) and subject to the proper choice of the generic function $\mathcal{C}_{\rm i}$. Here, we present two different kinds of the pressure profiles. The first one allows for the existence of the equatorial torus being determined by the solution 
\begin{eqnarray}
\label{sol+}
h^+&=&\frac{\Gamma-1}{\Gamma}H^+ +h^+_0\\
H^+&=&\frac{k_0}{\frac{4Q}{r}-2Br^2\omega\sin^2{\theta}}-\frac{1}{2}\ln{\left(2\mathcal{P}\right)},
\end{eqnarray}
corresponding to the chosen generic function 
\begin{eqnarray}
\label{generic+}
\mathcal{C}^+=k_0\left(\frac{2Q}{r}-Br^2\omega\sin^2{\theta}\right)^{-2},
\end{eqnarray}
with $h^+_0$ and $k_0$ being constants which have to be further specified. The other one embodies the polar clouds. These are determined by the solution
\begin{eqnarray}
\label{sol*}
h^*&=&\frac{\Gamma-1}{\Gamma}H^*+h^*_0\\
H^*&=&-\frac{k_0(\frac{2Q}{r}-Br^2\omega\sin^2{\theta})^2}{4Q}-\frac{1}{2}\ln{\left(2\mathcal{P}\right)},
\end{eqnarray}
corresponding to the chosen generic function 
\begin{eqnarray}
\label{generic*}
\mathcal{C}^*=k_0\left(\frac{2}{r}-\frac{Br^2\omega\sin^2{\theta}}{Q}\right),
\end{eqnarray}
with the constant of integration $h^*_0$.

Dropping now the symbols $^+$ and $^*$, the corresponding density, pressure and specific charge profiles are determined through the relations
\begin{eqnarray}
\label{densities}
\rho&=&\left(\frac{{\rm e}^{h}-1}{\kappa}\right)^\frac{1}{\Gamma-1},\\
\label{pressures}
p&=&\kappa\rho^{\Gamma},\\
\label{speccharges}
q_{\rm s}&=&K\left(1+\frac{p}{\rho}\right)= K{\rm e}^{h}
\end{eqnarray}
in both the cases. Total charges of the structures are determined by the common relativistic volume integral
\begin{eqnarray}
\label{totcharges}
\mathcal{Q}&=&\int_{\mathcal{V}}q{\rm d}\mathcal{V}=4\pi \int_{r_{\rm in}}^{r_{\rm out}}\hspace{-0.2cm}\int_{\theta_{\rho_0}}^{\frac{\pi}{2}} q_{\rm s}\rho\sqrt{-g}\,\rm{d}\theta\,\rm{d}r,
\end{eqnarray}
where $\theta_{\rho_0}$ is the function determining zero density surface given numerically from the equation $\rho=0$, $r_{\rm in}$ and $r_{\rm out}$ are positions of the inner and outer edges of the structures, and $g=-r^4\sin^2\theta$ is the determinant of the metric tensor according to equation (\ref{metric})

For the sake of completeness, we can be interested in the estimation of the central temperature of the torus and the strength of the magnetic field generated by the tori. To find an order of magnitude for the temperature $\tau$, we can assume that in the first approximation the
pressure is of thermal nature (electrostatic corrections and radiation pressure are neglected). Then, in physical units, $\hat{p}\approx\hat{\rho} \hat{k}_{\rm B} \hat{\tau}/\hat{m}_{\rm u}$, where $\hat{k}_{\rm B}$ is the Boltzmann constant and $\hat{m}_{\rm u}$ the atomic mass unit. Assuming the polytropic equation of state $\hat{p} = \hat{\kappa} \hat{\rho}^{\Gamma}$, we get 
\begin{eqnarray}
\label{temperatures}
\hat{\tau}\approx\frac{\hat{m}_{\rm u}\hat{p}}{\hat{k}_{\rm B}\hat{\rho}}=\frac{\hat{m}_{\rm u}}{\hat{k}_{\rm B}}\hat{\kappa}\hat{\rho}^{\Gamma-1}=\frac{\hat{m}_{\rm u}}{\hat{k}_{\rm B}}\,\hat{c}^2({\rm e}^h-1),
\end{eqnarray}
where $\hat{c}$ is the speed of light.

To become familiar with the final solutions for the pressure (density, charge density) profiles (\ref{pressures}) - (\ref{speccharges}), and to clearly describe their dependencies on the many parameters being involved, we assume the polytropic exponent $\Gamma=5/3$ being fixed (in general varying as $1<\Gamma<2$), i.e., not entering the following dependencies discussion. 
As we can see from relations (\ref{pressures}) - (\ref{temperatures}), all the tori characteristics are fully determined through the auxiliary $h$-function and the polytropic coefficient $\kappa$. Note that the correction function $K$ is not a free parameter here; it has been fixed through the generic function $C$, given by the integrability condition (\ref{condition1}), and uniquely determines the solutions (\ref{sol+}) and (\ref{sol*}). 

The $h$-function directly depends on the electromagnetic background parameters $B$ and $Q$, torus rotation $\omega$, matter charge parameter $k_0$ and on the integration constant $h_0$. For simplicity, these parameters are referred to as the $h$-parameters. As we can see from relations (\ref{pressures})-(\ref{speccharges}), reshaping the profiles of $h$-function through these parameters directly determines the space characteristics of the torus. In more details, through the $h$-parameters, we can set the $h$-zero surface (zero-density surface) corresponding to the surface of the torus, determining the volume (size) of the torus, its edges, etc. Looking at the form of the solution (\ref{sol+}) or (\ref{sol*}), we can clearly see that the position of the torus center (determined by the equations $\partial_r h=0$, $\partial_{\theta} h=0$) is specified through the electromagnetic parameters $B$, $Q$, $k_0$ and rotational parameter $\omega$ only. Reshaping the profiles of the $h$-function through the $h$-parameters also directly determines the pressure, density, charge density, specific charge and temperature profiles, and the total charge as well. 

As for the polytropic coefficient $\kappa$, contrary to the $h$-parameters, it does not influence the space characteristics of the torus at all. It also does not influence the specific charge and temperature distribution throughout the torus. From the mentioned characteristics, it only modifies pressure, density and charge density profiles, and the total charge. 

For comparison with the ambient dominant magnetic field, we can estimate the strength of the magnetic field generated by the rotating torus close to its, e.g., outer edge. Assuming the equatorial torus, resembling the mathematical regular torus of its cross section radius $(r_{\rm out}-r_{\rm in})/2$, shrinking it to the infinitely thin ring at $r=r_{\rm c}$, we estimate the magnetic field strength as 
\begin{eqnarray}
\label{torusfields} 
\mathcal{\hat{B}}\approx\frac{\hat{\mu}_0 \hat{I}}{2\pi(\hat{r}_{\rm out}-\hat{r}_{\rm c})},
\end{eqnarray}
in the distance $r_{\rm out}$ from this center, with $\hat{\mu_0}$ being the vacuum permeability and $\hat{I}=\hat{\mathcal{Q}\hat{\omega}}/{(2\pi)}$ the total current through the torus. More detailed discussion of the self-magnetic fields can be found, e.g., in paper \cite{Tur-Kol:2013} 
\section{Equatorial tori and polar clouds construction}
\begin{figure*}
\centering
\includegraphics[width=1\hsize]{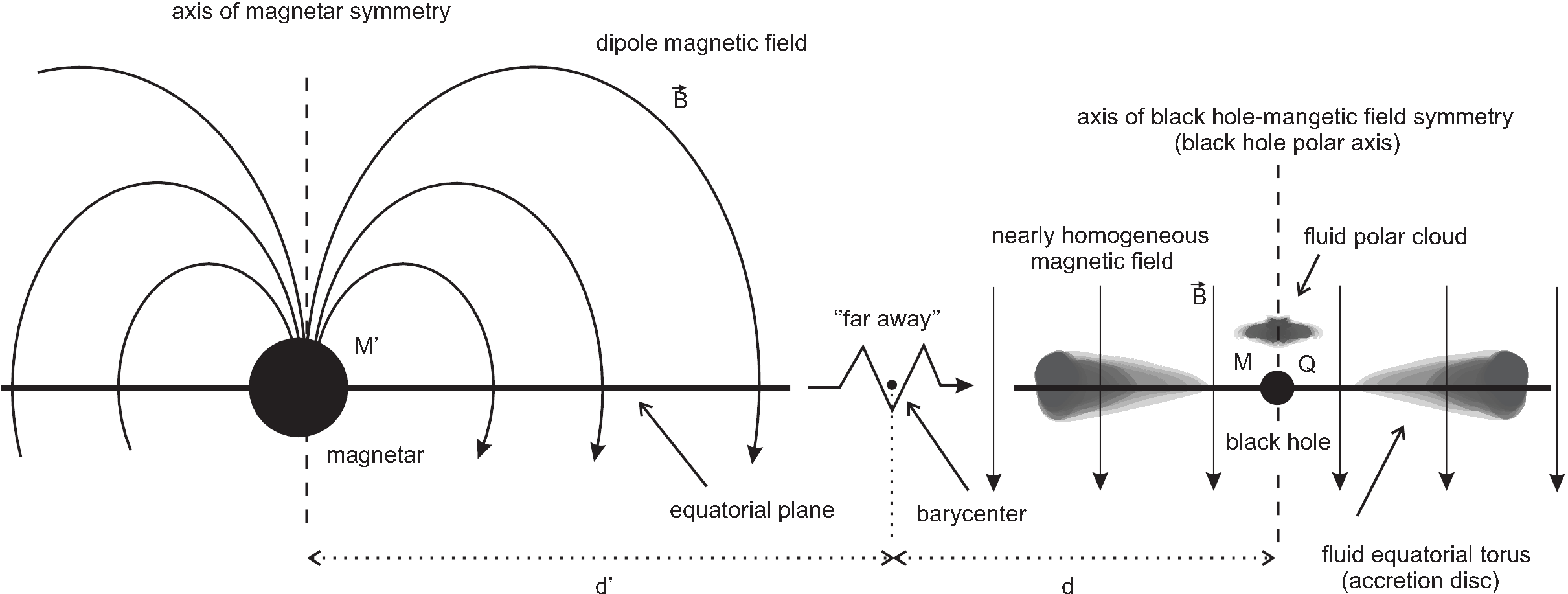}
\caption{Illustration of binary system of a magnetized star with mass $M'$ and a black hole with mass $M$ and charge $Q$ orbiting around a common center of mass in distances $d'$ and $d$. The dipole-type magnetic field $\vec{B}$ of a magnetized star appears to be almost uniform in close vicinity of a distant black-hole companion, which is encircled by a torus-like equatorial configuration and a polar cloud.}
\label{Fig:2}
\end{figure*}
Constructing the tori in the considered background of the Schwarzschild-Wald field, we can start with determination of the torus center $r_c$, specified through the parameters $B$, $Q$, $k_0$ and $\omega$. Although this work is intended theoretically, it is illustrative to chose these parameters within ranges of astrophysically realistic situation.
We should note that in our study the presence of the magnetic field $B$ and the black hole charge $Q$ plays crucial role for the existence of the charged tori, since the electromagnetic interaction with the charged tori must be strong enough to compete with the gravity. Moreover, because of the assumption of the very weak self-field of the rotating torus, we have to be aware of high values of the specific charge throughout the tori and must choose maximal possible astrophysical values for $B$ and $Q$. Otherwise, weak ambient electromagnetic field would require the tori with high specific charges, to preserve the strong interaction. Then, the tori would have to be very small or very diluted, in order to generate weak self-field. 

The galactic, nearly homogeneous, magnetic fields do not reach the required values, being weak as $\hat{B}\lesssim 10^{-4}\,{\rm T}$. They imply very high specific charges throughout the constructed tori, which is contrary to the assumption of negligible magnetic self-field generated by the rotating torus. Then, both the interacting external an self-electromagnetic fields have to be considered \cite{Cre:2011,Cre:2013,Cre-Kov:2013}. Another scenario having almost homogeneous magnetic field can be, however, related to a black hole of stellar mass $\hat{M}$ being a part of a binary system with a magnetar of mass $\hat{M}'$ equipped with a dipole type magnetic field of strength up to $\hat{B}\lesssim 10^{12}\,{\rm T}$ on its surface, the strongest magnetic field estimated to exist. While Fig. \ref{Fig:1} illustrated the idealized system of a magnetized black hole, Fig. \ref{Fig:2} provides a broader view of how such a system could form. Relatively far away from the magnetar, the field can be sufficiently well liken to homogeneous magnetic field for latitudes close to the equatorial plane. This assumption must be well met for the distance $(\hat{d}+\hat{d}') \gtrsim 10^3\,{\rm km}$ far from the magnetar, where $\hat{d}$ and $\hat{d}'$ are the distances of the black hole and magnetar from their barycentrum.

\subsection{Exemplary equatorial torus and polar cloud}
Adopting the common dipole magnetic field strength dependence $\sim(\hat{d}+\hat{d}')^{-3}$ and assuming the realistic radius of the magnetar  to be $\approx 10\,{\rm km}$, we get the possible magnetic field strength $\hat{B}\lesssim 10^6 {\rm T}$. In our consideration, we assume, for simplicity $\hat{M}=\hat{M}_{\odot}\doteq 1.99\times 10^{30}\,{\rm kg}$, which gives, in dimensionless units the limit $B\lesssim 10^{-10}$.

\subsubsection{Center position}
\begin{figure*}
\centering
\includegraphics[width=0.9\hsize]{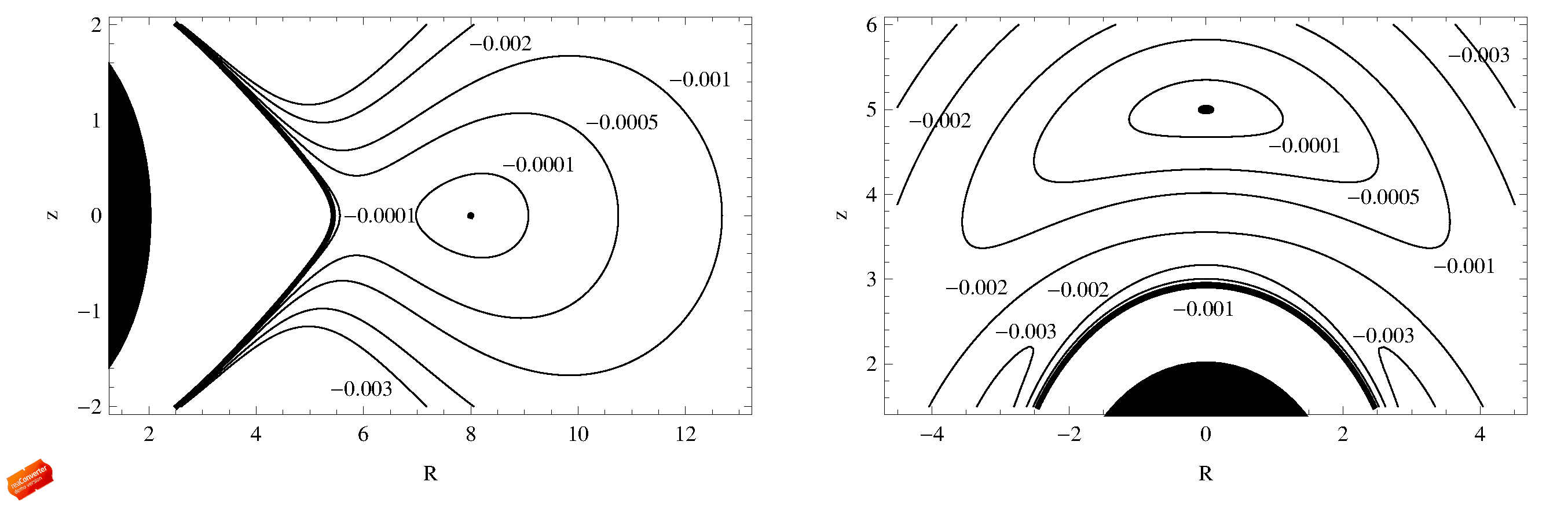}
\caption{Exemplary equatorial torus (left) and polar cloud (right) presented by means of poloidal sections of the function $h$, corresponding to the solutions (\ref{sol+}) and (\ref{sol*}) of equations (\ref{pressure2}). The thick curves denote zero-contours of the $h$-function, corresponding to the contours of zero equi-pressure and equi-density surfaces, bounding the tiny torus centered in the equatorial plane (left) and the polar cloud centered on the axis of symmetry. The equatorial torus solution, centered at $r_{\rm c}=8$ is characterized by the parameters $B\doteq 8.78\times 10^{-11}$, $Q=-10^{-11}$, $k_0=10^{-11}$  $\omega=1/100$ and $h_0=0.1134551345$ leading to relatively small torus with the edges in the equatorial plane at at $r_{\rm in}\doteq7.985$ and $r_{\rm out}\doteq8.015$. The polar cloud solution, centered at $r_{\rm c}=5$,  then corresponds to the parameters $B=10^{-11}$, $Q\doteq 4.17^{-11}$, $k_0=10^{11}$, $\omega=-1/1000$ and $h_0=0.10313150$, leading also to relatively small cloud with the edges in the axis of symmetry at $r_{\rm in}\doteq 4.976$ and $r_{\rm out}=5.024$. The matter characterized in both the cases by $\Gamma=5/3$. The related specific charge and density profiles are illustrated in Fig.\ref{Fig:4}.}
\label{Fig:3}
\end{figure*}

For an illustration of the situation in the case of equatorial torus (see left part of Fig. \ref{Fig:3}), we choose the center of the torus relatively close to the black hole at the radius $r_{\rm c}=8$. According to the condition $\mathcal{R}(r=r_c,\theta=\pi/2)=0$, rewritten in the form 
\begin{eqnarray}
\label{center_eq}
k_0=\frac{(Br_{\rm c}^3\omega-2Q)^2(1-r_{\rm c}^3\omega^2)}{r_{\rm c}(Br_{\rm c}^3\omega+Q)(r_{\rm c}-2-r_{\rm c}^3\omega^2)},
\end{eqnarray}
we get the magnetic field strength close to the mentioned limit $B\approx 10^{-10}$, for the black hole charge $Q=-10^{-11}$ (corresponding to $\bar{Q}=Q\bar{M}\doteq -1.48\times 10^{-8}\,{\rm m}$ and to $\hat{Q}=\bar{Q}\hat{c}^2(4\pi\hat{\epsilon}_0)^{1/2}\hat{G}^{-1/2}\doteq -1.71\times 10^{9}\,{\rm C}$), for $k_0=10^{-11}$ and for the angular velocity $\omega=1/100$ (corresponding to $\bar{\omega}=\omega/\bar{M}\doteq 6.67\times 10^{-6}\,{\rm m}^{-1}$ and $\hat{\omega}=\bar{\omega} \hat{c}\doteq 2.03\times 10^3\,{\rm rad.s}^{-1}$), where we chose as the representative value of the black hole mass $\bar{M}=\bar{M}_{\odot}=1.48\,{\rm km}$. The constants $\hat{G}$ and $\hat{\epsilon}_0$ are the gravitational constant and the vacuum permitivity expressed in physical units. Namely, we get $B\doteq 8.78\times 10^{-11}$, corresponding to $\bar{B}=B/\bar{M}\doteq 5.92\times 10^{-14}\,{\rm m}^{-1}$ and $\hat{B}=\bar{B}\hat{c}(4\pi\hat{\epsilon_0} \hat{G})^{-1/2}\doteq0.207\times 10^6\,{\rm T}$. 

Of course, as it follows from relation (\ref{center_eq}), the mentioned maximal value of the magnetic field strength can be reached for another combination of the parameters $Q$, $k_0$ and $\omega$ for the torus centered at $r_c=8$. However, these choices are also limited, as we explain later. For simplicity, in accordance with the possible configuration, we can afford to assume $M=M'=M_{\odot}$ and $d'=d$ in the binary system. For a magnetar having radius $R_{\rm m}=12\,{\rm km}$, which we consider as a representative value, we get the black hole at the distance $d+d'\doteq 1370$, immersed in the assumed magnetic field $B\doteq 8.78\times 10^{-11}$. Here, the reasonable Newtonian approximation gives the orbital period of the binary system and the corresponding azimuthal linear velocity of the black hole-torus system relatively to the barycentrum
\begin{eqnarray}
\hat{P}=2\pi\sqrt{\frac{(\hat{d}+\hat{d}')^{3}}{\hat{G}(\hat{M}+\hat{M}')}},\quad \hat{V}=\frac{2\pi\hat{d}}{\hat{P}}
\end{eqnarray}
so that $\hat{P}\doteq 1.11\,{\rm s}$ and $\hat{V}\doteq 0.02\,{\rm \hat{c}}$. Now, we have to chose the angular velocity of the torus $\omega$ sufficiently high, so that the orbital motion of the torus around the black hole is much faster than the orbital motion of the black hole around the barycenter to ensure quasi-static magnetic field, in which charged matter is circling around the black hole. This means that in the black hole system, the azimuthal velocity of, e.g., the center of the equatorial torus $\hat{v}=\omega \hat{c}\sqrt{g_{\phi\phi}}/\sqrt{-g_{tt}}\gg\hat{V}$. We return to this crucial limitation in section \ref{staticMF}. Since there is the limitation by the speed of light, we cannot fully meet this condition. We can, however, require $\hat{v}>\hat{V}$. For our choice $\omega=1/100$, we get $\hat{v}=0.08\,{\rm \hat{c}}$. Higher values of $\omega$ thus lead to ultrarelativistic motion, which we do not consider due to the initial assumptions. As it follows from relation (\ref{center_eq}), desiring to include the charge of the black hole into the electromagnetic interaction so that it plays nearly the same role as the magnetic field strength, it must be of order $10^{-11}$. The magnitude order of $k_0$ is then the same as the magnitude order of the magnetic field strength, i.e., $10^{-11}$. 

The similar reasoning can be realized in the case of polar clouds (see right part of Fig. \ref{Fig:3}). In the case of polar cloud, we chose its angular velocity $\omega=-1/1000$ (corresponding to $\bar{\omega}\doteq 6.67\times 10^{-7}\,{\rm m}^{-1}$ and to $\hat{\omega}\doteq 0.203\times 10^3\,{\rm rad.s}^{-1}$) and we chose the same representative values of the black hole mass $\bar M=\bar{M}_{\odot}$. The polar cloud is constructed in the magnetic field of the strength $B=10^{-11}$ (corresponding to $\bar{B}\doteq 6.67\times 10^{-15}\,{\rm m}^{-1}$ and $\hat{B}\doteq 23.6\times 10^3\,{\rm T}$) and centered at $r_{\rm c}=5$. Due to the relation  
\begin{eqnarray}
\label{center_pol}
k_0=\frac{- r_{\rm c}^2}{(r_{\rm c}-2)(Br_{\rm c}^3\omega-2Q)},
\end{eqnarray}
following from the condition $\mathcal{R}(r=r_c,\theta=0)=0$, the charge of the black hole is chosen to be $Q\doteq 4.17\times 10^{-11}$ (corresponding to $\bar{Q}\doteq 6.15\times 10^{-8}\,{\rm m}$ and $\hat{Q}\doteq 7.14\times 10^{9}\,{\rm C}$) and the scaling constant is set to be $k_0=10^{11}$. Here we get the distance between the magnetar and black hole-torus system $d+d'\doteq 2825$. The related orbital period and  azimuthal linear velocity of the black hole-torus system relatively to the barycentrum are $\hat{P}\doteq3.2\,{\rm s}$ and $\hat{V}\doteq 0.01\,{\rm \hat{c}}$. The azimuthal velocity of the outer edge of constructed polar cloud in the black hole system is $\hat{v}\doteq-0.0001\,{\rm \hat{c}}$, corresponding to the chosen $\omega=-1/1000$. 

\subsubsection{Characteristics profiles}
\begin{figure*}
\centering
\includegraphics[width=1\hsize]{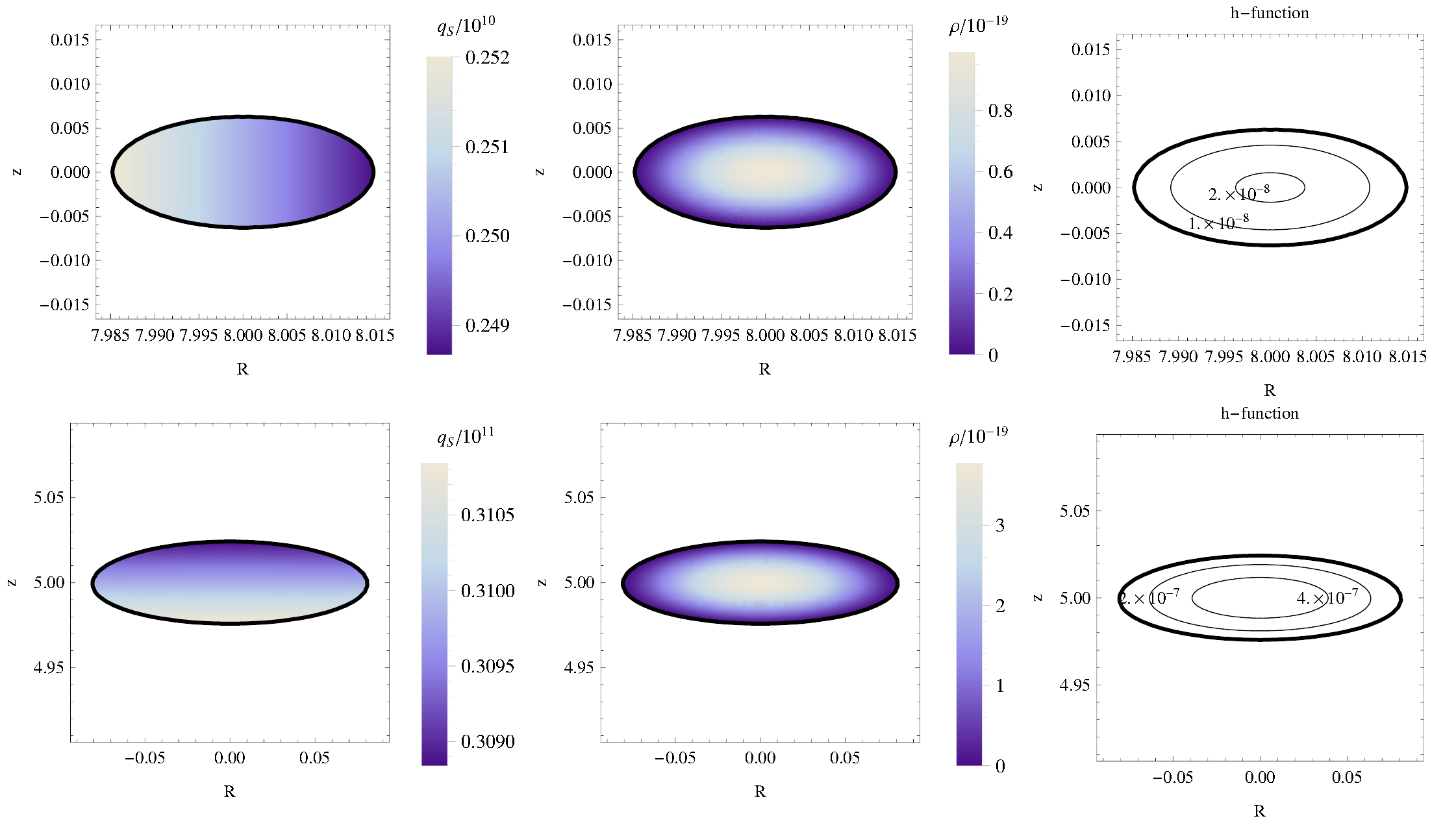}
\caption{Exemplary equatorial torus (top) and polar cloud (low) presented by means of specific charge $q_{\rm s}$ and density $\rho$ profiles determined by relations (\ref{speccharges}) and (\ref{densities}), corresponding to the equatorial solution (\ref{sol+}) and polar solution (\ref{sol*}) for the $h$-function. The corresponding $h$-function contours outside the torus and cloud, and parameters of the solution are given in Fig. \ref{Fig:3}; the $h$-function contours inside the structures are presented in the right column of this figure. The density reach its maximum $\rho_{\rm max}\approx 10^{-19}$ at the center of the equatorial torus $r_{\rm c}=8$ for the polytropic coefficient $\kappa=10^5$; the specific charge descents outwards from the black hole, being of the order of $10^9$. In the case of polar cloud, we have $\rho_{\rm max}\approx 10^{-19}$ in the center $r_{\rm c}=5$, for the chosen coefficient $\kappa=10^6$, and the specific charge being of the order $10^{10}$ throughout the polar cloud.}
\label{Fig:4}
\end{figure*}

Having the center of the torus and the polar cloud set (having $B$, $Q$, $k_0$, $\omega$ parameters fixed), it is now very important to realize that through the last $h$-parameter, i.e., through the additive constant of integration  $h_0$, we can set all the remaining tori space characteristic, the density, pressure, specific charge, charge density and temperature profiles, and the total charge, as mentioned above.
The $\kappa$ parameter influences the profiles of density, pressure, charge density and also the total charge.  
In our presented example of equatorial torus (see left part of Fig.\ref{Fig:4}), we chose $h_0=0.1134551345$ leading to relatively small torus with the edges in the equatorial plane at $r_{\rm in}\doteq7.985$ and $r_{\rm out}\doteq8.015$. According to relations (\ref{pressures}) - (\ref{torusfields}), the matter characterized by $\Gamma=5/3$ and $\kappa=10^{5}$ takes the maximal density $\rho_{\rm max}\approx 10^{-19}$ at the center of the torus (corresponding to $\hat{\rho}_{\rm max}\approx 10\,{\rm kg.m^{-3}}$), maximal pressure $p_{\rm max}\approx 10^{-27}$ (corresponding to $\hat{p}_{\rm max}\approx 10^{11}\,{\rm Pa}$), $q_{\rm s}\approx 10^9$ and the maximal temperature $\hat{\tau}_{\rm max}\approx 10^5\,{\rm K}$. The total charge of the rotating torus, $\mathcal{Q}\approx 10^{-12}$, being one order lower in comparison with the test charge of the black hole, gives rise to the poloidal-like magnetic field of the strength being roughly $\hat{\mathcal{B}}\approx 10^3\,{\rm T}$ at the edge of the torus. Such a field is sufficiently weak in accordance with fundamental assumption of our model, as it is two orders lower in comparison with the ambient field $\hat{B}$.  
In the case of polar cloud (see right part of Fig. \ref{Fig:4}), we chose $h_0=0.10313150$ leading also to relatively small cloud with the edges along the axis of symmetry at $r_{\rm in}\doteq 4.976$ and $r_{\rm out}=5.024$. The matter characterized by $\Gamma=5/3$ and $\kappa=10^{6}$ leads to $\rho_{\rm max}\approx 10^{-19}$ (corresponding to $\hat{\rho}_{\rm max}\approx 10^2\,{\rm kg.m^{-3}}$), $p_{\rm max}\approx 10^{-25}$ (corresponding to $\hat{p}_{\rm max}\approx 10^{13}\,{\rm Pa}$), $q_{\rm s}\approx 10^{10}$, total charge of the torus $\mathcal{Q}\approx 10^{-12}$ and the temperature $\hat{\tau}_{\rm max}\approx 10^6\,{\rm K}$.         

The chosen parameters $\kappa$ and $h_0$ in both cases presented above lead to relatively feasible pressure, density, specific charge and temperature profiles of the studied structures; the total charges of the structures are one order lower in comparison with the charges of considered black holes. In the case of the equatorial torus, we keep charge separation in the system - the total charge of the torus is of the opposite sign in comparison with the charge of the black hole. In the case of the polar cloud, this cannot be achieved, because of this unique configuration, which requires the additional repulsive force competing with gravity. By definition within the assumed model, the centers of equatorial tori and polar clouds correspond to the loci with the maximal pressure, density and temperature, matching the loci of $h$-function maxima according to relations (\ref{densities}), (\ref{pressures}), and (\ref{temperatures}). On the other hand, the maxima of charge density are shifted from the centers towards black holes, since the specific charge distributions are not constant, but descending outwards black holes, as determined by the choice of the generic functions (\ref{generic+}) and (\ref{generic*}).

In both presented cases, the charged structures are relatively tiny as compared to the size of the central black hole. As we have explained above, we can change the size of our toroidal structures with the fixed center by changing (increasing) the values of $h_0$. But this will lead to the increase of the density and consequently to the increase of the charge density and total charge, i.e., to the enhancement of the self-electromagnetic field, which would violate our fundamental assumption. We can then, however, decrease the density enough by increasing $\kappa$, to keep the same total charge (and the weak self-field); the temperature profile remains the same, being independent on such a change,  and increases only with increasing $h_0$.  

\subsection{Problem of static magnetic field and matching of torus characteristics} 
\label{staticMF}
As we mentioned above, although the magnetic field of the magnetar in the assumed distance $d+d'\approx 10^3$ can be clearly assumed to be homogeneous within the black hole-torus system, apparently, it cannot be assumed as static, since the linear velocities of the circling charged matter $v$ are not of higher orders in comparison with the circulation $V$ of the binary system. The velocity of the binary system in the examples presented above almost reaches the order of speed of light, so that it cannot be overcome much by speeding up the rotation of the torus. Clearly, the possible solution of this `non-static' field problem can be in slowing down the black hole-torus configuration in the binary system by increasing the size of the binary system $d+d'$ in many orders. Unfortunately, after that, the ambient magnetic field $B$, in which the torus will be immersed, simultaneously weakens. Because of this, assuming the torus center $r_{\rm c}$ at the same position relatively to the black hole, we have to dramatically increase the specific charge throughout the torus, which is necessary to keep the same electromagnetic interaction, since the gravity between the torus and black hole, and inertial (centrifugal) properties remain unchanged. Moreover, technically, 
according to relation (\ref{center_eq}) with changed $B$, we have to proportionally decrease $Q$ and increase $k_0$, keeping the same very high $\omega$, which is necessary for $v\gg V$. Having also the same size and density profiles of the torus, i.e., having the same values of the remaining $h_0$ and $\kappa$ parameters, this would lead to enormous increase of the charge density and total charge of the torus, i.e., also to strong electromagnetic self-field of the torus. 

Thus, the density and size of the torus must be appropriately modified by changing $\kappa$ and $h_0$. We demonstrate such modifications in the two following examples of charged equatorial torus circling around black hole in the binary system with magnetar in the distance $d+d'\doteq 13700$, i.e., ten times further than in the equatorial torus presented above, implying now the azimuthal velocity $\hat{V}\doteq 0.006\,{\rm \hat{c}}$, being nearly one order lower than in the previous case. In such a distance, the tori will be constructed in the magnetic field of the strength $B\doteq 8.78\times 10^{-14}$, corresponding to $\hat{B}\doteq0.207\times 10^3 {\rm T}$. 
For a comparison with the equatorial torus constructed before, we will assume the same position of the torus center at $r_{\rm c}=8$, the same angular velocity $\omega=1/100$ and polytropic exponent $\Gamma=5/3$. According to the relation (\ref{center_eq}), now, the charge of the black must be $Q=-10^{-14}$ and the scaling constant must be set to $k_0=10^{-14}$.     
\subsubsection{Density modification through $\kappa$-parameter} 
The problem of the large total charge of the torus can be solved by increasing the parameter $\kappa$ in order to decrease the density profiles and then, finally, to get reasonable total charge generating weak self-field. 

In the example of equatorial torus presented here, we consider the parameter $h_0$ being the same as in the previous case, i.e., $h_0=0.1134551345$. We also preserve the torus size and the edges in the equatorial plane at $r_{\rm in}\doteq7.985$ and $r_{\rm out}\doteq8.015$. To decrease the density, we chose $\kappa=10^{9}$, which gives $\rho_{\rm max}\approx 10^{-25}$ (corresponding to $\hat{\rho}_{\rm max}\approx 10^{-5}\,{\rm kg.m^{-3}}$), $p_{\rm max}\approx 10^{-33}$ (corresponding to $\hat{p}_{\rm max}\approx 10^{5}\,{\rm Pa}$) and $q_{\rm s}\approx 10^{12}$. The total charge of the torus is now $\mathcal{Q}\approx  10^{-15}$, originating the poloidal magnetic field of the strength $\hat{\mathcal{B}}\approx 1 {\rm T}$ close to the edge of the torus. The temperature profile remains the same as in the previous case of equatorial torus, i.e., $\hat{\tau}_{\rm max}\approx 10^5 {\rm K}$, being independent on $\kappa$. As we can see, the torus here is diluted enough, generating magnetic field by two orders lower in comparison with the weak ambient field from the magnetar considered here (see Fig. \ref{Fig:5}). 

\subsubsection{Size and density modification through $h_0$-parameter}
\begin{figure*}
\centering
\includegraphics[width=1\hsize]{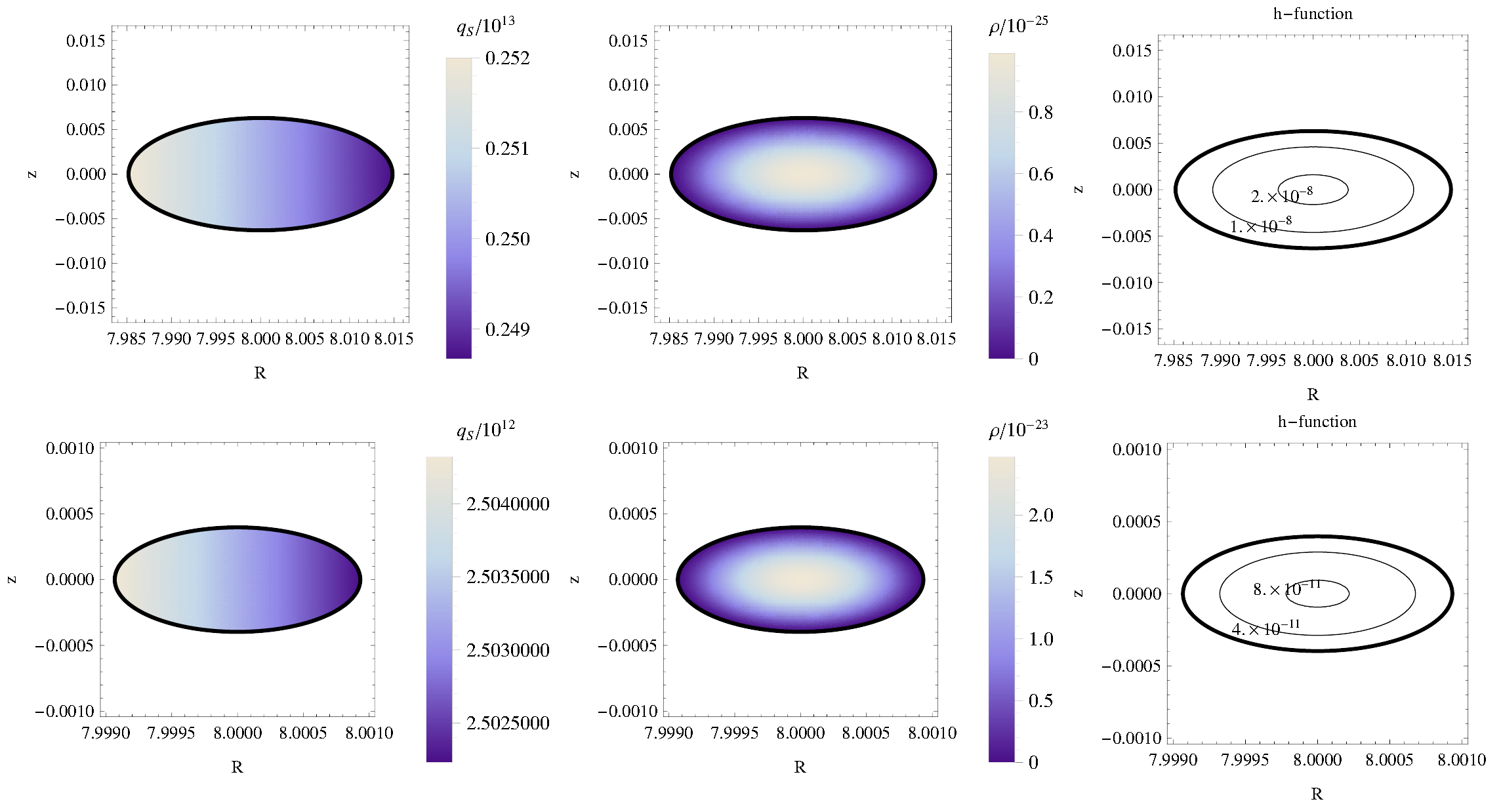}
\caption{Modifications of the exemplary equatorial torus presented in Fig. \ref{Fig:4} through the $\kappa$-parameter (top) and $h_0$-parameter (low). The torus is presented through specific charge $q_{\rm s}$ and density $\rho$ profiles determined by relations (\ref{speccharges}) and (\ref{densities}), corresponding to the equatorial solution (\ref{sol+}) for the $h$-function (with its contours presented in the right column of the figure), with the parameters $r_{\rm c}=8$, $B\doteq 8.78\times 10^{-14}$, $Q=10^{-14}$, $\omega=1/100$, $k_0=10^{-14}$ and $\Gamma=5/3$. The profiles illustrated in the top part of the figure correspond to $h_0=0.1134551345$ and $\kappa=10^{9}$, which yields the maximal density $\rho_{\rm max}\approx 10^{-25}$ in the center of the torus with the edges at $r_{\rm in}\doteq7.985$ and $r_{\rm out}\doteq8.015$, and the specific charge of the order $10^{12}$. The profiles illustrated in the low part of the figure correspond to $h_0=0.11345511319$ and $\kappa=10^{5}$, giving rise to the maximal density $\rho_{\rm max}\approx 10^{-23}$ in the center of the torus with the edges at $r_{\rm in}\doteq7.999$ and $r_{\rm out}=8.001$, and to the specific charge of the order $10^{12}$.}
\label{Fig:5}
\end{figure*}

To solve the problem with large total charge of the torus, we can also decide to decrease the size of the torus (having the same position of the center determined through $B$, $Q$, $k_0$ and $\omega$ parameters) by decreasing the parameter $h_0$, while keeping the parameter $\kappa$ unchanged. This primarily decrease the size of the torus, and also slightly the density profiles; consequently the total charge and the generated self-field can be neglected.   
We present here the situation with the polytropic coefficient $\kappa=10^5$ (the same as in the first presented case of equatorial torus) and with $h_0=0.11345511319$. Thus, we obtain torus smaller then in the previous case, the torus with the edges in the equatorial plane at $r_{\rm in}\doteq7.999$ and $r_{\rm out}=8.001$. Here, the maximal values of considered characteristics reach $\rho_{\rm max}\approx 10^{-23}$ (corresponding to $\hat{\rho}_{\rm max}\approx 10^{-2}\,{\rm kg.m^{-3}}$), $p_{\rm max}\approx 10^{-33}$ (corresponding to $\hat{p}_{\rm max}\approx 10^{5}\,{\rm Pa}$) and $q_{\rm s}\approx 10^{12}$. The total charge of the torus is the same as in the previous case $\mathcal{Q}\approx 10^{-15}$, generating the field of the strength $\hat{\mathcal{B}}\approx 10\,{\rm T}$ at the edge of the torus. As for the temperature, it decreases due to decreasing of $h$ parameter, now reaching maximal value $\hat{\tau}_{\rm max}\approx 10^3\,{\rm K}$.
Compared to the previous case, the torus here is not so diluted, but it is still small enough to generate magnetic field one order lower in comparison with the magnetic field of distant magnetar (see Fig. \ref{Fig:5}).\\

\subsubsection{Density limit and extension of the structures}
The question is, how rarefied matter we can afford in order to justify the used magneto-hydrodynamic approach. Usually, the limit for the kinetic approach applicability is the number density up to $10^{24}\,{\rm m^{-3}}$. Assuming particles of proton mass (specific charge $q_{\rm s}\approx 10^{18}$), we get the mass density limit $\hat{\rho}_{\rm MHD}\gtrapprox10^{-3}\,{\rm kg.m^{-3}}$. We, however, assume our fluid not to consist from proton plasma. The specific charge profiles in our tori (considered above and below) are several (from six to nine) orders lower. But this is an `average' specific charge of a bulk of particles, which can be presented like, e.g., a charge particle (of proton mass), bounded by $10^6-10^9$ neutral particles (of neutron mass). To support this scenario, the limit for applicability of our approach must be thus considered several orders higher than $\hat{\rho}_{\rm MHD}$. Clearly, the first two examples of charged structures, presented in the previous subsection, with the density maxima $10\,{\rm kg.m^{-3}}$ and $10^2\,{\rm kg.m^{-3}}$ safely satisfy this condition. On the other hand, the ones presented in this subsection, reaching only $10^{-5}\,{\rm kg.m^{-3}}$ and $10^{-2}\,{\rm kg.m^{-3}}$, are highly speculative. 

In all the four presented examples, we get tiny charged structures when related to the size of the central black hole. This is, however, in agreement with the conception of the rigid rotation of the tori, which can be kept only over short distances in real systems. On the other hand, in accordance with our assumption of negligible conductivity and rigid rotation, we would appreciate higher densities throughout the constructed structures. As it follows from the discussion above, this is highly conditioned by the maximal value of the external magnetic field in which our structures are immersed, i.e., the highest limit for the magnetic field strength of the magnetar. In our considerations we assumed the limit $\hat{B}\lesssim 10^{12}\,{\rm T}$ on its surface. On the stellar-scale, the relevant objects can be Anomalous X-ray pulsars, which exhibit episodic bright bursts and are thought to be young neutron stars with strong magnetic fields of the order of $\sim10^{11}\,{\rm T}$. It is expected that the activity is triggered by rearrangement of the magnetic field, whereas alternative explanations consider accretion of matter onto a neutron star with a less-extreme dipole magnetic field of $\sim10^8\,{\rm T}$, where the accreting matter originates from a fallback disc or a cloud around the neutron star when the magnetic field structure and intensity change.

As mentioned above, the assumptions adopted in our approach are suitable for low-conductivity medium consisting of electrically charged particles 
(such as dust particles) that interact via gravitational as well as electromagnetic forces acting on them from the surrounding environment and the nearby compact star. In this context, fallback accretion discs are expected to form from the remnant material after a supernova \cite{Alp:2013}. These have properties similar to debris discs or asteroid belts and they are supposed to encircle the neutron star and it can even trigger its collapse to the black hole when the critical mass is exceeded.
Because of significant content of dust and relatively low temperature in fallback discs, they are suitable candidates for the structures explored in the present paper. They should exhibit themselves by the dust emission in mid- and far-infrared spectral bands, and in millimeter radio 
wavelengths \cite{Wan:2006}. Therefore, by the mass-scalling properties of accreting compact objects, similar physics can be applied 
to vastly different types of object.

\section{Conclusions}
In this paper, we studied electrically charged toroidal structures encircling a charged black hole immersed into an asymptotically homogeneous magnetic field. In contrast to the preceding work, considering charged tori around pure charged black hole, here, the addition of the ambient magnetic field enriches the background for the studied structures with possible astrophysical interpretation.
Moreover, we imposed the assumption of rigid (constant angular velocity) rotation of the structures, instead of the previously used constant angular momentum distribution of the rotating matter. We focused on regular toroidal structures centered in the equatorial plane and also on unique structures, referred to as polar clouds, located around polar axis of the black hole, which is parallel to magnetic field lines. 

The purpose of this paper is to prove the theoretical existence of such structures, where the presence of electric charge is the necessary condition. We have shown that the charged tori could exist, having relatively feasible physical characteristics, like pressure, density, charge density and temperature profiles. For this purpose, we fixed the background and torus parameters so that the characteristics of the modeled structures do not deviate from the generally accepted limits for realistic matter existence and, simultaneously, do not violate the fundamental assumptions of the model - the neglect of the generated self-fields. 

We discussed here only four configurations in details. But the free parameters in the model enable construction of very wide spectrum of toroidal configurations, which can be investigated henceforward. There is also a challenge to take the spin of the black hole into account, for now being considered zero. Higher values of this spin parameter can substantially change the range of achieved profiles of characteristics throughout the tori. However, contrary to the fully analytical study applied here, such an investigation would require a numerical approach. 

We found that the density, pressure and temperature profiles, extension of the constructed tori and polar clouds, and other characteristics highly depend on the strength of the external magnetic field, into which such structures are immersed. As we have shown, the strongest magnetic fields in the universe, the dipole-like fields in the vicinity of magnetized neutron stars, provide more or less sufficient fields, even in the distances far from neutron star, where the magnetic fields can be considered as homogeneous, over small regions close to the equatorial planes. However, the equatorial tori and polar clouds constructed there, orbiting a black hole companion of the magnetar, are close to their low density limit. More realistic density profiles would require stronger magnetic fields, which are present closer to the magnetized star, where the black hole, as the magnetic star companion, cannot be located because of the unrealistic binary system characteristic. This gives rise to the question, whether the rigidly rotating equatorial tori and polar clouds can exist directly around the magnetized neutron star in its close vicinity, in the strong dipole-type magnetic field. Detailed study of this kind is outlined for the future work. Another line of future investigations concerns the interacting external and self-electromagnetic fields, under the framework developed in papers \cite{Cre:2011,Cre:2013}.

\appendix*
\section{Pressure equations derivation}
\label{appendix}
Within general relativity, motion of charged perfect fluid is generally described by conservation laws and 
Maxwell's equations
 \begin{eqnarray}
\label{cons}
\nabla_{\beta}T^{\alpha\beta}&=&0,\\
\label{Maxw}
\nabla_{\beta}F^{\alpha\beta}&=&4\pi J^{\alpha},\\
\label{Maxw1}
\nabla_{(\gamma}F_{\alpha\beta)}&=&0,
\end{eqnarray}
where the \mbox{4-current} density $J^{\alpha}$ can be expressed in terms of the charge density $q$, electrical conductivity $\sigma$ and 
\mbox{4-velocity} $U^{\alpha}$ of the fluid through Ohm's law
\begin{eqnarray}
\label{Ohm}
J^{\alpha}=q U^{\alpha}+\sigma F^{\alpha\beta}U_{\beta}.
\end{eqnarray}
The electromagnetic tensor $F_{\alpha\beta}=\nabla_{\alpha}A_{\beta}-\nabla_{\beta}A_{\alpha}$ here, being given in terms
of the vector potential $A_{\alpha}$, describes the vacuum external electromagnetic field which comes through the fluid and also the internal 
electromagnetic self field of the fluid, and can be written as
 \begin{eqnarray}
F^{\alpha\beta}=F^{\alpha\beta}_{\rm EXT}+F^{\alpha\beta}_{\rm INT}.
\end{eqnarray}
Also the stress-energy tensor $T^{\alpha\beta}$ can be split into two parts; the matter and  
electromagnetic parts as
\begin{eqnarray}
\label{T}
T^{\alpha\beta}=T^{\alpha\beta}_{\rm MAT}+T^{\alpha\beta}_{\rm EM},
\end{eqnarray}
where
\begin{eqnarray}
\label{Tmat}
T^{\alpha\beta}_{\rm MAT}&=&(\epsilon+p)U^{\alpha}U^{\beta}+pg^{\alpha\beta},\\
\label{Tem}
T^{\alpha\beta}_{\rm EM}&=&\frac{1}{4\pi}\left(F^{\alpha}_{\;\;\gamma}F^{\beta\gamma}-\frac{1}{4}F_{\gamma\delta}F^{\gamma\delta}g^{\alpha\beta}\right).
\end{eqnarray}

From the conservation law (\ref{cons}), by using the stress-energy tensor decomposition (\ref{T}) and, being aware that for electromagnetic field, due to Maxwell equations (\ref{Maxw}) and  (\ref{Maxw1}), it holds \cite{MTW}
\begin{eqnarray}
\nabla_{\beta}T^{\alpha\beta}_{\rm EM}=-F^{\alpha\beta}J_{\beta},
\end{eqnarray}
we get the equation 
\begin{eqnarray}
\label{premaster}
\nabla_{\beta}T^{\alpha\beta}_{\rm MAT}=F^{\alpha\beta}J_{\beta},
\end{eqnarray}
where $J_{\beta}$ is the \mbox{4-current} density generated by the charged matter. 

However, since the basic assumption of our model is to consider only the tori being composed from the `test matter' (from gravitational and also from electromagnetic point of view), we do not take into account an impact of electromagnetic field generated by this \mbox{4-current}. It means, we have  $F^{\alpha\beta}_{\rm INT}\ll F^{\alpha\beta}_{\rm EXT}$ and can write $F^{\alpha\beta}=F^{\alpha\beta}_{\rm EXT}$. Then we obtain from equation (\ref{premaster}) the final formula  
\begin{eqnarray}
\label{master}
\nabla_{\beta}T^{\alpha\beta}_{\rm MAT}=F^{\alpha\beta}_{\rm EXT}J_{\beta}.
\end{eqnarray}

Our model describes the case of charged matter moving in the azimuthal direction only, i.e., having the \mbox{4-velocity} of the form $U^{\alpha}=(U^t,U^{\phi},0,0)$ (which is equivalent to the condition of zero conductivity throughout the torus, $\sigma=0$), thus, the \mbox{4-current} $J^{\alpha}$ given by the Ohm law (\ref{Ohm}) must have the only non-vanishing spatial component   
\begin{eqnarray}
J^{\phi}=q U^{\phi}.
\end{eqnarray}
Assuming now an axially symmetric and stationary background with the axis of symmetry aligned with the torus symmetry axis, from formula (\ref{master}), we directly obtain two non-linear partial differential equations (\ref{pressure}) for the pressure $p$ throughout the torus.
Finally, note that since the electromagnetic field $F^{\alpha\beta}_{\rm EXT}$ is prescribed in our scenario, we do not solve Maxwell's equations (\ref{Maxw}) here. More details related to the model of charged perfect fluid can be found in the paper \cite{Kov-etal:2011}.

\begin{acknowledgments}
The Institute of Physics and Astronomical Institute have been operated 
under the project ``Albert Einstein Center for Gravitation and Astrophysics'', Czech Science Foundation GA\v{C}R No. 14-37086G. The authors JK, PS, and ZS would also like to express their acknowledgment for the Institutional support of Faculty of Philosophy and Science, Silesian University in Opava; CC acknowledges financial support by the Italian Foundation ``Angelo Della Riccia'' (Firenze, Italy) and the research project of the Czech Science Foundation GA\v{C}R grant No. 14-07753P. We also acknowledge the COST Action MP1304 ``Exploring fundamental physics with compact stars''. 
\end{acknowledgments}

\providecommand{\noopsort}[1]{}\providecommand{\singleletter}[1]{#1}%

\end{document}